\documentclass[reprint,
groupedaddress,
showpacs,preprintnumbers,
 amsmath,amssymb,
 aps,
prl
]{revtex4-1}
\usepackage{graphicx}
\usepackage{dcolumn}
\usepackage{bm}
\usepackage{mathrsfs}

\begin{document}

\title{Crossed surface flat bands of Weyl semimetal superconductors}

\author{Bo Lu}
\affiliation{Department of Applied Physics, Nagoya University, Nagoya 464-8603, Japan}

\author{Keiji Yada}
\affiliation{Department of Applied Physics, Nagoya University, Nagoya 464-8603, Japan}

\author{Masatoshi Sato}
\affiliation{Department of Applied Physics, Nagoya University, Nagoya 464-8603, Japan}

\author{Yukio Tanaka}
\affiliation{Department of Applied Physics, Nagoya University, Nagoya 464-8603, Japan}

\date{\today}

\begin{abstract}
It has been noted that certain surfaces of Weyl semimetals have
 bound states forming open Fermi arcs, which are never seen in typical
 metallic states.
We show that the Fermi arcs enable them to support an even more exotic
surface state with {\it crossed} flat bands in the superconducting state.
We clarify the topological origin of the crossed flat bands and the relevant
 symmetry that stabilizes the cross point. We also
discuss their possible experimental verification by tunneling spectroscopy.
\end{abstract}

\pacs{74.50.+r, 73.20.-r, 74.20.Rp, 03.65.Vf}

\maketitle

\affiliation{Department of Applied Physics, Nagoya University, Nagoya
464-8603, Japan}

\affiliation{Department of Applied Physics, Nagoya University, Nagoya
464-8603, Japan}

\affiliation{Department of Applied Physics, Nagoya University, Nagoya
464-8603, Japan}


\textit{Introduction.}--
Weyl semimetals (WSMs) are
three-dimensional materials that support pairs of bulk gapless
points that are effectively described by Weyl fermions \cite{Murakami07,wan11,Fang11,
Burkov11prl,Kim12,Chen12,Magnetdoping,Hosur13,Morimoto2014}.
The characteristic
band-touching points may be viewed as topological magnetic
monopoles in momentum space,
which predicts many interesting phenomena such as anomalous
Hall effects, chiral
anomalies \cite{Adler1969,Bell1969,NN1983,Aji241101,PhysRevB.86.115133,PhysRevB.87.235306}, and
magneto-electric effects \cite
{Landsteiner1,Landsteiner2,Wang,Oshikawa,Chen}.
Candidate materials include pyrochlore iridates \cite{wan11},
 HgCr$_{2}$Se$_{4}$ \cite{Fang11,Fang2012} and magnetically doped Bi$_{2}$Se$_{3}$ family \cite{Magnetdoping,Nomura2014}.
A simpler realization in
a topological insulator multilayer has also been
proposed \cite{Burkov11prl,Burkov11prb}.
The most fundamental and striking prediction for
WSMs is the existence of Fermi {\it arcs} on their boundary \cite{wan11,haldane}.
Whereas ordinary electrons in metals form closed Fermi surfaces in
momentum space, the surface bound states of WSMs form open
arcs at the Fermi energy, not closed circles.
The Fermi arcs reported in experimental studies of
high-$T_{\rm c}$ cuprate superconductors were not true
arcs, but the suppression of the density of states due to the formation of pseudogap states \cite{Damascelli}.
In contrast,
WSMs host true arcs that are
terminated by the projection of band-touching Weyl points on the surface
Brillouin zone (BZ).
Such exotic states may carry topological flows, so they
provide the aforementioned variety of nontrivial phenomena in low-energy
physics.
WSMs require breaking time-reversal or inversion symmetry
\cite{Burkov11prl}.
Below we consider time-reversal breaking WSMs.

Upon slight doping, WSMs have disconnected Fermi surfaces,
each of which surrounds one of the band-touching Weyl points.
It has been studied that
either spacial uniform
(e.g., BCS s-wave) or nonuniform (e.g., Fulde--Ferrell--Larkin--Ovchinnikov (FFLO) state) Cooper pairing
can be formed on these Fermi surfaces
\cite{Moore12,Gilbert13,Aji14,Aji2014}.
Interestingly,
the uniform superconducting state is found to support
bulk gap nodes on the Fermi surface even for a constant
$s$-wave pairing \cite{Moore12}.

In this letter, we show that the nodal superconducting WSMs may
support even more exotic surface bound states than Fermi arcs:
The novel surface states form {\it crossed} flat
bands, not simple arcs.
Like Weyl points,
nodes (antinodes) in superconducting states generally have positive (negative) monopole
charges in momentum space.
Thus, there are topological flows from nodes (or to antinodes),
and corresponding surface zero energy flat bands are produced.
We find that in the nodal superconducting state in WSMs,
each Fermi surface supports only nodes (or only antinodes).
Consequently, there arises a topological twist in the dispersion of the
surface Andreev bound states (SABSs), which make it possible to realize
such a complicated surface band structure.

\begin{figure}[htbp]
 \begin{center}
  \includegraphics*[width = 70 mm]{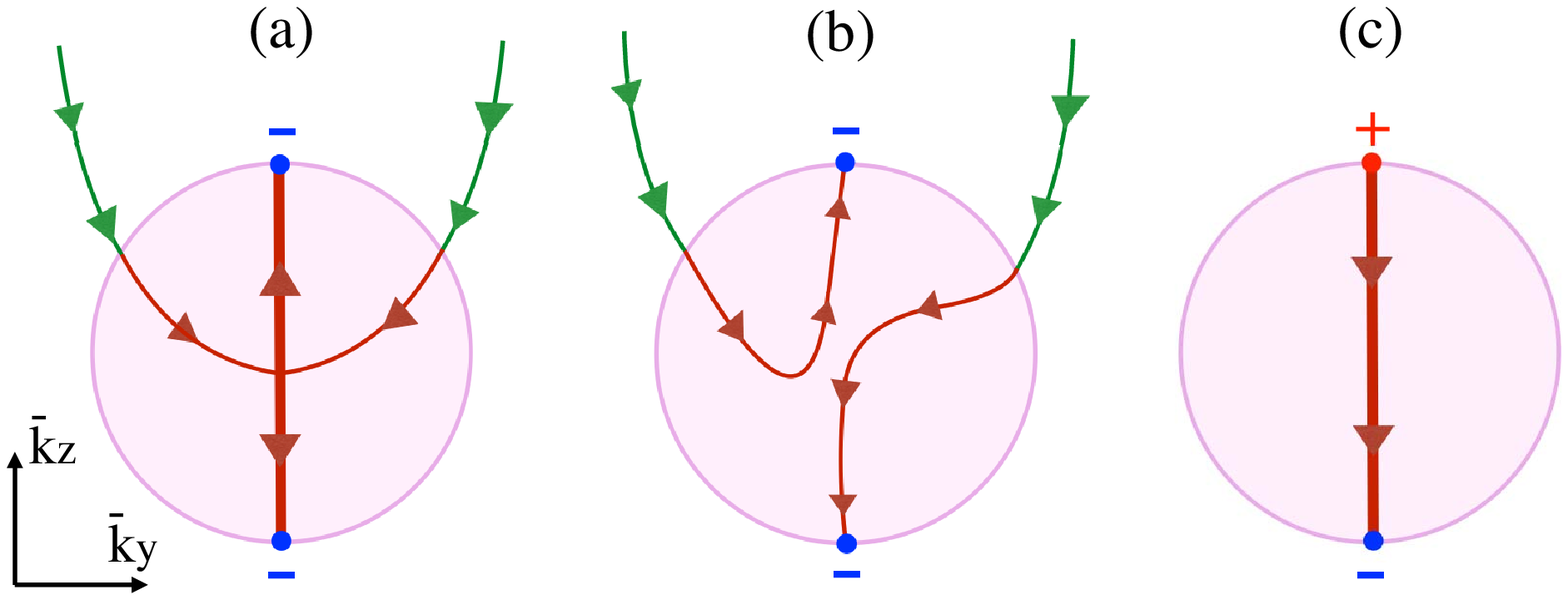}
 \end{center}
 \caption{(color online). Schematic illustration of surface flat bands
in doped superconducting WSMs (a) with and (b) without magnetic mirror reflection symmetry,
and (c) that for $^3$He-A phase.
The shaded circle (dots)
indicates the Fermi surface (point nodes) projected on the surface BZ.
The inserted signs at point nodes correspond to the signs of monopole charge.
Arrows show the topological flows and are regarded as the zero energy flat bands.
Simultaneously, the directions of the arrows show the chirality of the surface states:
The zero energy flat band has a rightward group velocity when facing the direction of the arrows.
}
 \label{fig1}
\end{figure}

Figure \ref{fig1}(a) illustrates how such a topological twist occurs.
A key ingredient is surface Fermi arcs in WSMs:
The Fermi arcs in time-reversal breaking WSMs does not have a Kramers partner,
which is necessary to form its BCS Cooper pair.
Thus, the Fermi arcs remain as surface gapless modes even in the superconducting state.
In the Nambu representation, we also have surface Fermi arcs of holes
as well as those of electrons.
Near the projected Fermi surface on the surface BZ, they merge into the zero energy SABS,
as illustrated in Fig.\ref{fig1}(a).
Preserving the topological flow denoted by arrows,
these zero energy flat bands bend at a crossing point and they are terminated at the projection of antinodes.
Note that a similar merging of surface states has been reported for
superconducting topological insulators \cite{hsieh,Yamakage,JPSJ.83},
although the relevant topological number and obtained spectrum are
completely different.
Later, on the basis of symmetry and topology, we will argue that the
existence of the crossing point of the flat bands
is rather general.
Such crossed band structures strongly enhance the surface density of
states, which might induce nontrivial low-energy phenomena.
Our findings also extend the possibility of controlling zero energy flat bands \cite{liunature04}.

In the following, using a concrete model of WSMs,
we demonstrate crossed surface flat bands in an $s$-wave
superconducting state.
Then, we identify the topological number responsible for the exotic
band structure and embody the topological arguments given above.
We will also discuss the experimental signatures.
We use $\hbar=1$ units and take the lattice spacing $a$
as $a=1$.


\textit{Model.}--As a model of WSMs with an $s$-wave pairing, we use the
two-band Hamiltonian
$
H=\mathrm{\frac{1}{2}}\sum\nolimits_{\bm{k}}\hat{c}_{\bm{k}}^{\dag
 }\mathcal{H}_{\bm{k}}\hat{c}_{\bm{k}}
$ \cite{RanY11, Moore12}
with
\begin{eqnarray}
&&\mathcal{H}_{\bm{k}}=t\sin k_{x}\sigma _{y}\tau _{z}-t\sin k_{y}\sigma _{x}\tau
_{0}+(t_{z}\cos k_{z}-M)\sigma _{z}\tau _{z}  \notag \\
&&+m(2-\cos k_{x}-\cos k_{y})\sigma _{z}\tau _{z}-\mu \sigma _{0}\tau
_{z}-\Delta \sigma _{y}\tau _{y},
\label{eq2}
\end{eqnarray}%
where the spinor $\hat{c}_{k}$ is given by $(c_{k\uparrow },c_{k\downarrow
},c_{-k\uparrow }^{\dag },c_{-k\downarrow }^{\dag })^{T}$,
$\sigma _{i}\left( \tau _{i}\right) $ is the Pauli matrix
in spin (Nambu) space, and $t$ $(t_{z})$ is the hopping in the
$k_{x}$--$k_{y}$ plane (along the $k_{z}$-axis).
$M(\equiv t_{z}\cos Q)$ denotes a magnetic order or a Zeeman field that
breaks the time-reversal symmetry, $\mu$ is the chemical potential, and
$\Delta$ is a conventional $s$-wave pair potential.
We also introduce a parameter $m$ to control the
position of the Weyl points.
When $m=0$, the normal state ($\Delta=0$) possesses
four pairs of bands touching Weyl points at
$\left( 0,0,\pm Q\right) $,
$\left( \pi ,0,\pm Q\right) $, $\left( 0,\pi ,\pm Q\right) $, and $\left( \pi
,\pi ,\pm Q\right)$, respectively.
Hence, upon slight doping with small $\mu$, we have eight
disconnected Fermi surfaces, each of which surrounds a Weyl point.
When $m$ is turned on, however, the latter three pairs of Weyl points located on the
BZ boundary move; for large $m$, they pair-annihilate.
Correspondingly, only two disconnected Fermi surfaces surrounding
Weyl points at $\left( 0,0,\pm Q\right)$ survive.



Each Fermi surface of the doped WSM, in contrast to those of ordinary metals, does
not have spin degeneracy because of time-reversal breaking and strong
spin--orbit interaction.
The spin and momentum are locked on the Fermi surface, so a spinless system is realized effectively.
In this situation, even an $s$-wave pairing may host topological
superconductivity \cite{Sato03,Fu2008,STF09,Alicea10,Jay,lutchyn10,oreg10}.
In the present case, the simplest choice of the
pair potential generates point nodes of the superconducting gap
at the north and south poles of the Fermi surfaces \cite{Moore12}.




\textit{Surface Andreev bound states.}--
Surface Andreev bound states (SABSs) \cite{ABS,Hara,Hu,Kashiwaya} are a powerful
probe of topological superconductivity because they directly reflect
bulk topological structures \cite{Schnyder08,Flat11,tanaka12}.
To identify the nodal topological
superconductivity in doped WSMs,  we now examine the SABSs.
Using an efficient way to calculate the lattice Green's function \cite{PhysRevB.55.5266},
we can obtain
the SABSs
from the poles of the Green's function.
We choose the surface perpendicular to the $x$ direction.

\begin{figure}[htbp]
 \begin{center}
  \includegraphics*[width = 85 mm]{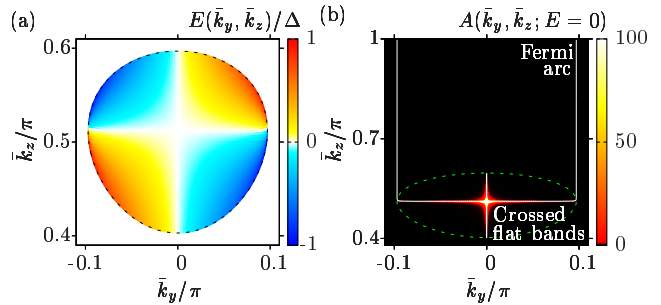}
 \end{center}
\caption{(color online)
(a) Energy dispersion $E\left( \bar{k}_{y},\bar{k}%
_{z}\right)$ of SABS inside the projected Fermi surface near $(0,0,Q)$.
(b) Quasiparticle spectra function at zero energy.
Dashed line denotes the projected Fermi surface.
$t=t_{z}=1$, $Q=\pi /2
$, $\mu =0.3$, $\Delta =0.001$, and $m=0.8$ are chosen for both (a) and (b).
}
 \label{fig2}
\end{figure}

Figure \ref{fig2}(a) shows the obtained SABSs.
The model parameters are chosen so that the WSM
has only two disconnected Fermi surfaces in the normal state.
We show only the SABS in the upper half-plane $(\bar{k}_z>0)$ of the
surface BZ, but a similar SABS exists in the lower half-plane
$(\bar{k}_z<0)$.
Clearly, Fig.\ref{fig2}(a) indicates that the SABS hosts a twisted
energy dispersion with
two crossed flat bands extended in the vertical and horizontal directions, respectively.
To be specific, consider the spectrum in the $\bar{k}_y$ direction.
For a fixed $\bar{k}_z$, the SABS appears as a chiral edge mode with a
linear dispersion,
$E=v(\bar{k}_z)\bar{k_y}$, for small $\bar{k}_y$.
The group velocity
$v(\bar{k_z})$ becomes zero at the position of the horizontal flat band,
and the sign of $v(\bar{k}_z)$ is reversed
when the chiral mode crosses the
horizontal flat band.
We also find that the horizontal flat band eventually becomes Fermi arcs, as illustrated  in  Fig.\ref{fig2}(b).
%
%

The crossed flat band structure can also be confirmed by
quasiclassical analysis. Consider a semi-infinite
superconducting WSM placed on the right ($x>0$) with a
semi-infinite insulator on the left ($x<0$). This can
be done by replacing the parameters $\Delta $ and $M$ in Eq. (\ref{eq2})
with $\Delta \Theta (x)$ and $t_{z}\cos Q+M_{0}\Theta (-x)$, respectively.
A large $M_{0}$ is
chosen so that the left side does not have Weyl cones and
thus become insulating. 
For weak pairing $0<\Delta \ll \mu $, we can use the quasiclassical BdG
Hamiltonian,
\begin{eqnarray}
\mathcal{H}& \mathrm{=}&it_{z}\sin Q\partial _{z}\sigma _{z}\tau
_{z}-it(\partial _{x}\sigma _{y}\tau _{z}-\partial _{y}\sigma _{x}\tau
_{0})-\mu \sigma _{0}\tau _{z}  \notag \\
&&-\Delta \Theta (x)\sigma _{y}\tau _{y}-M_{0}\Theta (-x)\sigma _{z}\tau
_{z},
\end{eqnarray}%
near the Weyl point at $\left( 0,0,Q\right) $. The solution of the BdG
equation $\mathcal{H}\Psi (r)=E\Psi (r)$ is given by $\Psi \left( r\right)
=e^{i\bar{k}_{y}y+i(\bar{k}_{z}-Q)z}[\Psi _{\rm I}(x)\Theta (-x)+\Psi
_{\rm S}(x)\Theta (x)]$ with
\begin{eqnarray}
&&\Psi _{\rm I}(x) =[s_{e}\Psi _{e1}+s_{h}\Psi _{h1}]e^{\kappa x},
\nonumber\\
&&\Psi _{\rm S}(x) =t_{e}\Psi _{et}e^{ik_{x}^{e}x}+t_{h}\Psi
_{ht}e^{-ik_{x}^{h}x}.
\end{eqnarray}%
Here $k_{x}^{e\left( h\right) }=\sqrt{%
q_{1}^{2}-\bar{k}_{y}^{2}-(q_{1}/q_2)^{2}\left( \bar{k}_{z}-Q\right) ^{2}%
}\pm i\zeta $ and
$\kappa t\mathrm{=}\sqrt{[(t_{z}\sin Q)(\bar{k}_{z}-Q)+M_{0}]^{2}+t^{2}\bar{k%
}_{y}^{2}-\mu ^{2}}$,
with $q_{1}=\mu /t$, $q_{2}=\mu /(t_{z}\sin Q)$ and
$\zeta =\frac{\sqrt{\Delta ^{2}\sin ^{2}\beta -E^{2}}}{t\sin \beta \cos \phi
}$.
With the parametrization of
$\bar{k}_{y}=q_{1}\sin \beta \sin \phi $
and $\bar{k}%
_{z}=q_{2}\cos \beta +Q$,
%
%
the four component amplitudes
are given by $\Psi
_{e1}=[t(\kappa +\bar{k}_{y}),-\eta ,0,0]^{T}$, $\Psi _{h1}=[0,0,t(-\kappa +%
\bar{k}_{y}),\eta ]^{T}$, $\Psi _{et}=[\gamma \tan (\beta /2),i\gamma
e^{i\phi },-ie^{i\phi }\tan (\beta /2),1]^{T}$, and $\Psi _{ht}=[\tan (\beta
/2),-ie^{-i\phi },i\gamma e^{-i\phi }\tan (\beta /2),\gamma ]^{T}$, where $
\eta =M_{0}+\mu +(t_{z}\sin Q)(\bar{k}_{z}-Q)$, and $\gamma =\frac{\sqrt{E+%
\sqrt{E^{2}-\Delta ^{2}\sin ^{2}\beta }}}{\sqrt{E-\sqrt{E^{2}-\Delta
^{2}\sin ^{2}\beta }}}$.
The coefficients $(s_{e(h)}, t_{e(h)})$ and the
energy $E$ are determined so as to satisfy the boundary condition $\Psi
_{I}(0)=\Psi _{S}(0)$. Then, we obtain the energy dispersion
as
\begin{equation}
E\mathrm{=}\Delta \bar{k}_{y}(\bar{k}_{z}-Q)/(q_{2}\sqrt{q_{1}^{2}-\bar{k}%
_{y}^{2}}),  \label{abs}
\end{equation}%
which clearly shows two flat bands along
$\bar{k}_{y}=0$ and $\bar{k}_{z}=Q$, respectively.
We also find that
the group velocity $v(\bar{k}_z)=\partial E/\partial
\bar{k}_y$ reverses its sign at $\bar{k}_{z}=Q$, as expected.

\begin{figure}[htbp]
 \begin{center}
  \includegraphics*[width = 80 mm]{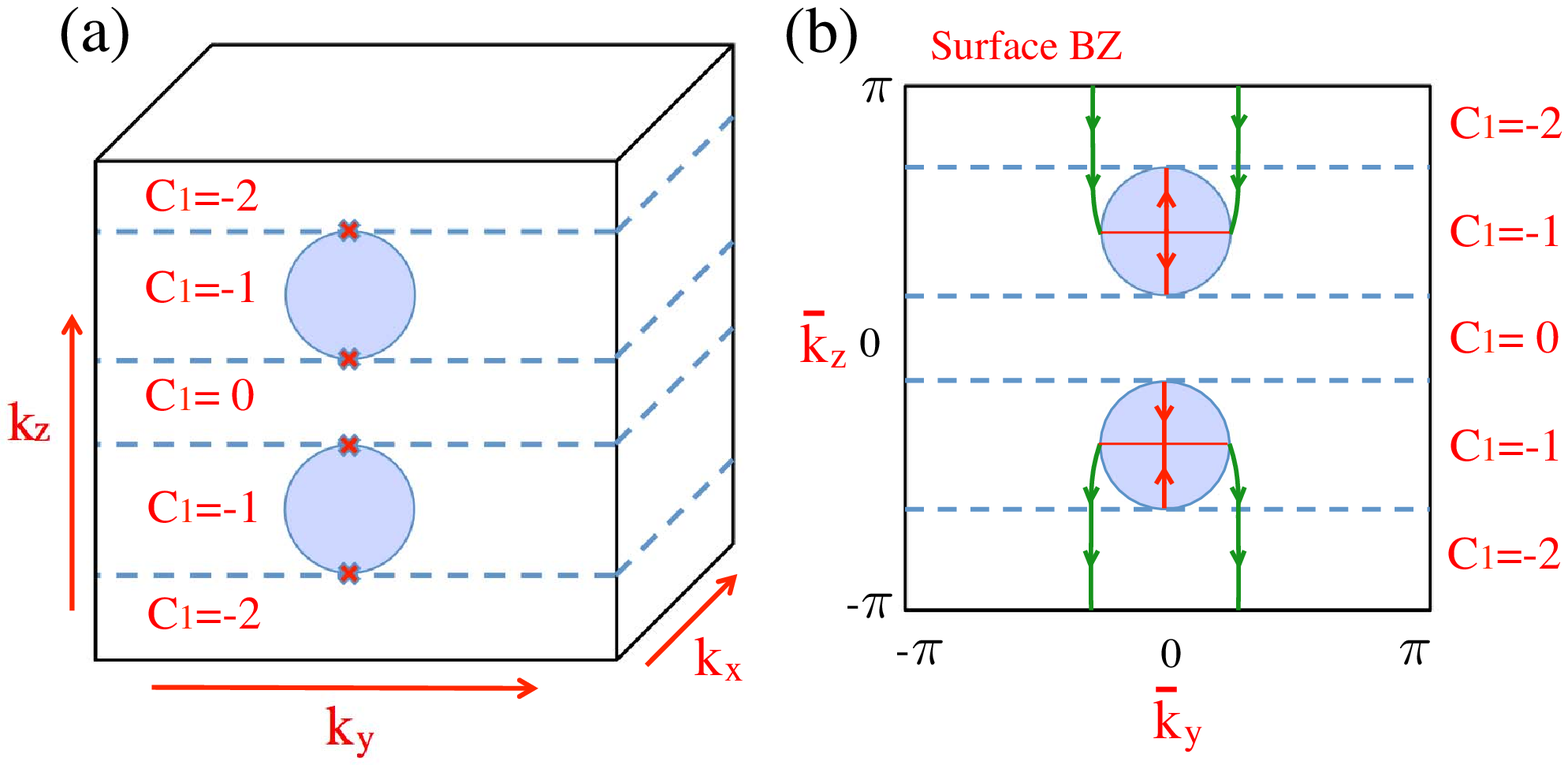}
 \end{center}
 \caption{(color online)
(a) Projection of Fermi surfaces (dark region) on $\overline{k}_{y}-\overline{k}_{z}$ surface BZ
and Chern number at a given $k_{z}$ in the BZ.
The Chern number $C_{1}(k_z)$ changes when $k_z$ crosses a plane
including a point node (red cross).
The amount of change is given by the monopole charge of the point node.
(b) Flat bands of SABS and Fermi arcs (lines with arrows) in the surface BZ.
%
}
 \label{fig3}
\end{figure}


\textit{Topological Analysis.}--
Now we would like to identify the bulk topology relevant to the crossed
surface flat bands, which reveals that the unusual band structure is
rather general for nodal superconducting states of WSMs.
For simplicity, we consider the simplest case with two disconnected Fermi
surfaces separated in the $k_z$ direction, but more complicated
cases can be discussed similarly.

As mentioned above, point nodes in superconducting WSMs behave like
monopoles in momentum space. Each point node is a source or sink of
the flux of the U(1) gauge field,
$
{\cal A}({\bm k})=i\sum_{E_n<0}\langle u_n({\bm k})|\nabla_{\bm
 k}|u_n({\bm k})\rangle,
$
where $|u_n({\bm k})\rangle$ is a bulk occupied state of the BdG Hamiltonian,
and the summation is taken for all occupied states.
To capture the topological structure, consider a plane $S$ that is
normal to the $k_z$-axis in the BZ.
If $S$ does not contain any point nodes, the total flux (over $2\pi$)
penetrating $S$ defines the first Chern number,
$
C_1(k_z)=\frac{1}{2\pi}\int_{S}d^2k \left[
 \nabla_{\bm k}\times {\cal A}({\bm k})
\right]_{z}
$,
where  $k_z$ is the position of $S$.
The Chern number is a topological invariant that remains the same
unless $S$ touches a gap-closing point.
When $S$ crosses a point node, however, the Chern number changes.
The change is equal to the total flux
leaving the point node; thus, it provides its monopole charge.

When $S$ is not close to the Fermi surface, the Chern number can be
evaluated rather easily.
In this case, we can turn off $\Delta$ without gap closing.
Therefore, the Chern number is essentially the same as that in the
normal state, though we have to take into account the contribution from
holes as well as electrons.
For inversion-symmetric WSMs,
the hole and electron contributions are found to be the same, so
the Chern number is doubled.
For instance, if the Chern number of electrons is -1 (0) when the
projection of $S$ on the surface BZ
crosses (does not cross) a Fermi arc,
in the superconducting state it becomes -2 (remains 0) if $S$ does
not overlap the Fermi surface.

This simple calculation gives an alternative
explanation of why the Fermi arcs remain in the
superconducting state. From the surface--boundary correspondence, a
nonzero bulk Chern number in WSMs ensures the existence of a Fermi arc;
thus, if the Chern number is doubled, the number of
Fermi arcs is also doubled by adding those of holes.
The resulting Fermi arcs remain gapless even in the superconducting state.

Interestingly, the same calculation also explains why point
nodes arise in the superconducting state.
Because each Fermi surface surrounds a Weyl point, at least
two Fermi arcs (i.e., those of holes and electrons) enter
the projection of the Fermi surface on the surface BZ.
From the above calculation, the Chern number corresponding to the
remaining Fermi
arcs should be even and nonzero, so a net flux of the U(1) gauge field
${\cal A}({\bm k})$ also enters the Fermi surface.
Therefore, flux conservation implies that there must be a
source or sink of flux near the Fermi surface.
In the normal state, the Weyl point is exactly the required source or
sink, but in the superconducting state, it cannot be because it is below
(or above)
the Fermi level in the doped WSMs,
so the Weyl point can no longer provide a monopole charge.
Alternatively, we must have an even number of superconducting gap nodes
on the Fermi surface, which supply a nonzero total monopole
charge. Figure \ref{fig3}
shows the Chern numbers and
corresponding SABSs in our model; the expected topological
structures are indeed realized.


Now we would like to explain the topological stability of the crossed
structure of the surface flat band:
Although time-reversal symmetry is broken in WSMs,
a combination of time-reversal and mirror reflection, which we call
as magnetic mirror reflection, can be preserved.
Actually, as well as the present model, candidate materials of WSMs such
as pyrochlore iridates, HgCr$_{2}$Se$_{4}$, and a topological insulator
multilayer
retain such magnetic mirror reflection symmetry.
Then if the surface of a WSM also keeps magnetic mirror reflection symmetry,
the surface flat bands should be symmetric under the reflection.
This symmetry forbids the cross point to be resolved as Fig.\ref{fig1}(b).
We can also assign the relevant topological invariant.
By combining the magnetic mirror reflection with the particle-hole
symmetry in the superconducting state,  we can introduce chiral operator
$\Gamma$ which anticommutes with the Hamiltonian on the
mirror plane.
For example, the Hamiltonian in Eq. (\ref{eq2}) anticommutes with
$\Gamma=\tau_x$ at $k_y=0$ and $\pi$.
The chiral operator enables us to define the winding number \cite{footnote}.
For the present model, we find that the winding number is nonzero when
$k_z$ is on the vertical flat band at $k_y=0$, while it changes the value at
point nodes and disappears outside the Fermi surface.
Thus, the vertical zero energy flat band at $\bar{k}_{y}=0$
is topologically protected, although it connects nodes (or antinodes)
with the same charge.
Consequently, the cross point cannot be resolved in order to keep
consistent topological flows.

\begin{figure}[htbp]
 \begin{center}
  \includegraphics*[width = 84 mm]{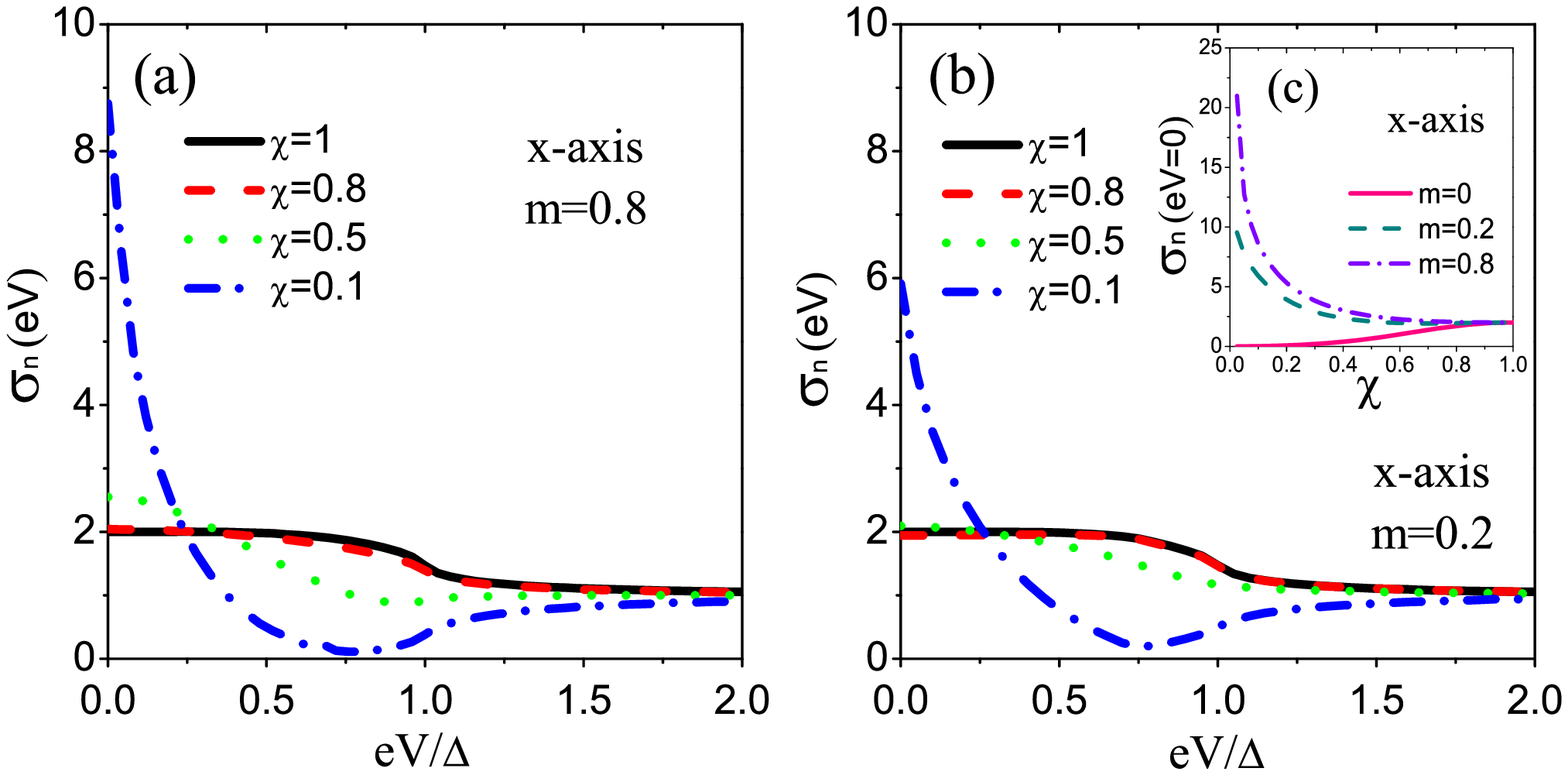}
 \end{center}
 \caption{(color online). Normalized tunneling conductance
 as a function of bias
voltage $\left( eV/\Delta \right) $ for $m$ equal to $\left( a\right) $ $0.8$ and $\left( b\right) $ $0.2$.
$\left(
c\right) $ Relation between height of normalized ZBCP and $%
\chi $. Other parameters are as the same as in Fig.\ref{fig2}.}
 \label{fig4}
\end{figure}

\textit{Experimental Signatures.}--
To explore the experimental signatures of the newly discovered SABSs, we
consider a normal metal/superconductor (NS) junction in a doped WSM and
calculate the normalized tunneling conductance $\sigma _{\rm n}(eV)=\sigma
_{\rm S}(eV)/\sigma _{\rm N}(eV)$ using the tight binding model in Eq.(\ref{eq2}) with an appropriate
boundary condition \cite{Kroemer,footnote2}. Here $\sigma
_{\rm S}(eV)$ is the conductance of the NS junction, and $\sigma _{\rm N}(eV)$ is that in
the normal state.
We denote the transmissivity at the interface by $\chi $,
where $\chi =0$ ($\chi =1$) corresponds to the low transparent limit (full
transmissivity)\cite{footnote2}.

Figures \ref{fig4}(a) and (b)
show the tunneling conductance $\sigma _{\rm n}$ in
the $x$ direction.
The obtained $\sigma_{\rm n}$
shows zero-bias conductance peaks (ZBCPs), where
the peak height is inversely proportional to $%
\chi $, as shown in Fig.\ref{fig4}(c).
Such features are rather similar to those of $d$-wave supercondcutors with line nodes \cite{TK95,Kashiwaya,Lofwander,Wong},
but not those of Sr$_2$RuO$_4$, which is a superconducting analog of the $^3$He-A
phase \cite{PhysRevB.56.7847} or a ferromagnet/superconductor junction on the surface
of a topological insulator \cite{Tanaka09}.
Although the chiral $p$-wave state supports a vertical
flat band, such a huge $\sigma_n$ at $eV=0$ cannot be obtained.
In contrast,  the crossing point of the flat bands in the superconducting
WSMs forms a saddle point in
the energy dispersion, producing
a Van Hove singularity in the surface density of states.
As a result, the ZBCP is very prominent.


\textit{Summary and Discussions.}--So far we have reported that WSMs with
surface Fermi arcs may support even more exotic crossed surface flat
bands in the superconducting state.
We found that the nontrivial topology of their normal state
results in bulk point nodes in the $s$-wave pairing state, which enables us
to design such nontrivial structure in condensed matter physics.
We have also revealed that the crossed flat band structure is
protected by magnetic mirror reflection symmetry.
Conversely, the latter result implies that we can control the crossed
flat bands by the perturbations which break the symmetry, e.g.,
the magnetic field along the $y$-axis.

Here we would like to compare our system with superfluid $^3$He-A phase.
Whereas superconducting WSMs support bulk point nodes similar to the A-phase
bulk \cite{Moore12}, we have found an important topological difference:
The $^3$He-A phase supports a node-antinode pair
with opposite monopole charge in the Fermi surface, and thus, without
topological twist,
a surface flat band starts at the projection of a node on the surface BZ and
ends at that of an antinode \cite{Volbook,Volovik2011}, as shown in Fig.\ref{fig1}(c).
In contrast,  the superconducting WSMs have nodes with the same charge,
so the twisting of surface bands may occur.

In closing, we remark on the possible generalization of this work.
Our topological arguments require three
conditions.
The first is a uniform pairing state.
A nonuniform pairing state such as the FFLO state mixes the Chern numbers
with different momenta, so it may destroy the topology of WSMs relevant
to the crossed flat bands.
The second is broken time-reversal symmetry in WSMs.
For time-reversal-symmetric WSMs, the Chern numbers of electrons and of holes are
canceled, so superconducting states cannot dominate the nontrivial
topology of WSMs.
The final condition is magnetic mirror reflection symmetry, which
stabilizes the crossed flat bands.
Once these three conditions are met,
we may have crossed surface flat bands for any pairing state.
Although an unconventional Cooper pair may create additional nodes on the Fermi surface,
the total number of monopole charges should be nonzero,
which allows the hosting of such complicated flat bands.

We thank A. Yamakage for useful discussion. This work was supported in part by Grants-in-Aid for
Scientific Research from the Ministry of Education, Culture,
Sports, Science and Technology of Japan (``Topological Quantum
Phenomena'' No. 22103005 and No. 25287085)
and by the EU-Japan program ``IRON
SEA.''


\pagebreak
\widetext
\begin{center}
\textbf{\large Supplementary Materials}
\end{center}
\setcounter{equation}{0}
\setcounter{figure}{0}
\setcounter{table}{0}
\setcounter{page}{1}
\makeatletter
\renewcommand{\theequation}{S\arabic{equation}}
\renewcommand{\thefigure}{S\arabic{figure}}
\renewcommand{\bibnumfmt}[1]{[S#1]}
\renewcommand{\citenumfont}[1]{S#1}

\vspace{3ex}
\noindent
\textbf{S1. Magnetic mirror reflection symmetry}

Here we discuss magnetic mirror reflection symmetry in WSMs.
The magnetic mirror reflection symmetry is a combined symmetry of mirror
reflection and time-reversal.
As discussed below, many WSMs naturally have mirror reflection symmetry
in a certain direction.

First, it should be noted that broken time-reversal is necessary to obtain a
nonzero Chern number $C_1(k_z)$ in inversion symmetric WSMs.
%
At the same time, we would like to point out that for the
nonzero Chern number $C_1(k_z)$, reflection symmetries in
directions normal to the $z$-axis must be broken.
Because the flux $[\nabla_{\bm k}\times {\cal A}({\bm k})]_z$ is odd
under time-reversal and the reflection, each of these symmetries forces
the integral $C_1(k_z)=\frac{1}{2\pi}\int d^2k[\nabla_{\bm k}\times
{\cal A}({\bm k})]_z$ to be zero.

Although each of them must be broken, their combined
symmetry is consistent with inversion symmetric WSMs.
Actually, we find that WSMs in pyrochlore iridates and HgCr$_2$Se$_4$
retain such magnetic mirror reflection symmetry.
In both WSMs, the magnetic mirror reflection with respect to the $(110)$
plane is preserved.
This fact could be physically understood as a consequence that magnetic
orders of these materials, which are necessary for nonzero $C_1(k_z)$,
keep symmetry of the original materials as much as possible.
We also find that a WSM in a topological insulator multilayer proposed
in Ref.[\onlinecite{S_Burkov11prl}] has a similar magnetic mirror reflection symmetry.
Therefore, if Cooper pairs do not break the magnetic reflection
symmetry, then the corresponding superconducting phase also retains the
magnetic reflection symmetry.

Now we argue the magnetic mirror reflection symmetry in our model,
\begin{eqnarray}
\mathcal{H}_{\mathbf{k}} &=&t\sin k_{x}\sigma _{y}\tau _{z}-t\sin
k_{y}\sigma _{x}\tau _{0}+(t_{z}\cos k_{z}-M)\sigma _{z}\tau _{z}
\nonumber \\
&&+m(2-\cos k_{x}-\cos k_{y})\sigma _{z}\tau _{z}-\mu \sigma _{0}\tau
_{z}-\Delta \sigma _{y}\tau _{y}.
\label{hami}
\end{eqnarray}%
The time-reversal operator is given by %
\begin{equation}
T=-i\sigma _{y}\mathcal{K}\tau _{0}.
\end{equation}%
with complex conjugation operator $\mathcal{K}$,
and the mirror reflection operator is given by
\begin{equation}
M_{xz}=i\sigma _{y}\tau _{0},
\end{equation}%
and each of these symmetries is broken in our model,
\begin{equation}
T{\cal H}_{\left(k_{x},k_{y},k_{z}\right) }T^{-1}\neq{\cal H}_{\left(-
k_{x},-k_{y},-k_{z}\right) },
\quad
M_{xz}{\cal H}_{\left( k_{x},-k_{y},k_{z}\right) }M_{xz}^{-1}\neq{\cal H}_{\left(
k_{x},k_{y},k_{z}\right) }.
\end{equation}%
However, combining these two operators $T$ and $M_{xz}$,
we can define time-reversal like operator %
$\tilde{T}=M_{xz}T$, which satisfies
\begin{equation}
\tilde{T}\mathcal{H}_{\left( k_{x},k_{y},k_{z}\right) }\tilde{T}^{-1}=%
\mathcal{H}_{\left( -k_{x},k_{y},-k_{z}\right) }.
\end{equation}%
This is the magnetic mirror reflection symmetry.
Combining with the particle-hole symmetry of the BdG Hamiltonian,
\begin{equation}
C\mathcal{H}_{\left( -k_{x},-k_{y},-k_{z}\right) }C^{-1}=-\mathcal{H}%
_{\left( k_{x},k_{y},k_{z}\right) },
\end{equation}%
with
\begin{equation}
C=\sigma _{0}\mathcal{K}\tau _{x},
\end{equation}%
we also obtain the {\it mirror chiral symmetry}
\begin{equation}
\Gamma \mathcal{H}_{\left( k_{x},k_{y},k_{z}\right) }\Gamma ^{-1}=-\mathcal{H%
}_{\left( k_{x},-k_{y},k_{z}\right) }.
\label{eq:mchiral}
\end{equation}%
with%
\begin{equation}
\Gamma =C\tilde{T}
\end{equation}%

This symmetry stabilizes the crossed surface flat bands as illustrated in
Fig.1(b) in the main text:
Equation (\ref{eq:mchiral}) implies that the flat bands with zero energy
should be symmetric under $k_y\rightarrow -k_y$.
Therefore, a problematic reconnection process in Fig.1(c) in the main text never happens as far as the magnetic mirror reflection symmetry ( and the
resultant magnetic chiral symmetry) is present.




\vspace{3ex}
\noindent
\textbf{S2. Winding number}

In this section, we show that the vertical flat band in Fig.\ref{fig2}
in the main text has a non-trivial topological number defined by
mirror chiral symmetry in Eq.(\ref{eq:mchiral}).
At $k_y=0, \pi$, the mirror chiral symmetry reduces to
\begin{equation}
\{\Gamma ,\mathcal{H}_{\left( k_{x},k_{y},k_{z}\right) }|_{k_{y}=0,\pi }\}=0,
\end{equation}%
so we can define the following one-dimensional winding number for
fixed $k_z$\cite{S_Fujimoto09},
\begin{eqnarray}
W= -\frac{1}{4\pi i}\int_{-\pi}^{\pi} dk_x {\rm tr}
\left[\Gamma {\cal H}^{-1}_{\bm k}\partial_{k_x}{\cal H}_{\bm k}
\right]_{k_y=0, \pi}
\end{eqnarray}
Following the discussion in Ref.[\onlinecite{S_Yada2011}], we can transform $%
\mathcal{H}_{\mathbf{k}}$ into anti-diagonalized form by unitary matrix $U$:
\begin{equation}
U\mathcal{H}_{\mathbf{k}}U^{\dag }=
\left(
\begin{array}{cc}
0 & q\left( \mathbf{k}\right)  \\
q^{\dag }\left( \mathbf{k}\right)  & 0%
\end{array}%
\right),
\end{equation}
with
\begin{equation}
q\left( \mathbf{k}\right) =\left(
\begin{array}{cc}
-\mu +D\left( \mathbf{k}\right)  & -\Delta -it\sin k_{x} \\
\Delta +it\sin k_{x} & -\mu -D\left( \mathbf{k}\right)
\end{array}%
\right)
\end{equation}
and
with $D\left( \mathbf{k}\right) =t_{z}\cos k_{z}-M+m(2-\cos
k_{y}-\cos k_{x})$. Using a parameter $\theta =\arg (\det q(\mathbf{k}))$,
the winding number $W$ at $k_{y}=0$ or $\pi $ can be evaluated as
defined as \cite{S_Flat11}:
\begin{equation}
W=\frac{1}{2\pi }\int_{-\pi }^{\pi }\frac{\partial \theta }{\partial k_{x}}%
dk_{x}.  \label{winding}
\end{equation}%
After a straightforward calculation, we obtain for $\Delta >0$%
\begin{eqnarray}
W_{k_{y}=0}\left( k_{z}\right)  &=&\frac{1}{2}[\mathrm{sgn}\left( \omega
_{1}\right) -\mathrm{sgn}\left( \omega _{2}\right) ], \\
W_{k_{y}=\pi }\left( k_{z}\right)  &=&\frac{1}{2}[\mathrm{sgn}\left( \omega
_{2}\right) -\mathrm{sgn}\left( \omega _{3}\right) ],
\end{eqnarray}%
where $\omega _{1}=\Delta ^{2}+\mu ^{2}-(t_{z}\cos k_{z}-M)^{2}$, $\omega
_{2}=\Delta ^{2}+\mu ^{2}-(t_{z}\cos k_{z}-M+2m)^{2}$ and $\omega
_{3}=\Delta ^{2}+\mu ^{2}-(t_{z}\cos k_{z}-M+4m)^{2}$.
In Fig.3(b)in the main text, we find that the winding number is nonzero when
$k_z$ is on
the vertical flat band at $k_y=0$.  The winding number disappears at
point nodes and it remains zero
outside the Fermi surface.
A more complicated case is also illustrated in Fig.3(d) in the main text.

Similar type of
flat bands of SABS as Majorana fermions is discussed in non-centrosymmetric
superconductors \cite{S_Tewari13,S_SauMA2012,S_Schnyder14,S_YSTY10,S_Mizuno} and
spin-orbit coupled systems\cite%
{S_STF09,S_Alicea10,S_Jay,S_Fujimoto09,S_oreg10,S_lutchyn10}.



\vspace{3ex}
\noindent
\textbf{S3. Tunneling conductance}

To calculate the tunneling conductance of NS (normal metal / superconductor)
junction based on the tight-binding model in Eqs. (1) and (2) in the main
text, we express the Hamiltonian $H$ in a lattice space, which is given by
\begin{eqnarray}
H &=&\sum\nolimits_{ijn}it(-\bar{c}_{i,j,n}^{\dag }\sigma _{y}\bar{c}%
_{i+1,j,n}+\bar{c}_{i,j,n}^{\dag }\sigma _{y}\bar{c}_{i-1,j,n}+\bar{c}%
_{i,j,n}^{\dag }\sigma _{x}\bar{c}_{i,j+1,n}-\bar{c}_{i,j,n}^{\dag }\sigma
_{x}\bar{c}_{i,j-1,n})/2 \nonumber \\
&&+t_{z}(\bar{c}_{i,j,n}^{\dag }\sigma _{z}\bar{c}_{i,j,n+1}+\bar{c}%
_{i,j,n}^{\dag }\sigma _{z}\bar{c}_{i,j,n-1})/2+\left( 2m-t_{z}\cos Q\right)
\bar{c}_{i,j,n}^{\dag }\sigma _{z}\bar{c}_{i,j,n} \nonumber \\
&&-m(\bar{c}_{i,j,n}^{\dag }\sigma _{z}\bar{c}_{i+1,j,n}+\bar{c}%
_{i,j,n}^{\dag }\sigma _{z}\bar{c}_{i-1,j,n}+\bar{c}_{i,j,n}^{\dag }\sigma
_{z}\bar{c}_{i,j+1,n}+\bar{c}_{i,j,n}^{\dag }\sigma _{z}\bar{c}_{i,j-1,n})/2
\nonumber \\
&&-\mu \bar{c}_{i,j,n}^{\dag }\bar{c}_{i,j,n}+\Delta c_{i,j,n\uparrow
}^{\dag }c_{i,j,n\downarrow }^{\dag }+\Delta c_{i,j,n\downarrow
}c_{i,j,n\uparrow ,}
\end{eqnarray}%
with $\bar{c}_{ijn}
=\left( c_{ijn\uparrow },c_{ijn\downarrow }\right) ^{T}$.
Here, $i$, $j$, and $n$ denote the site indexes in $x$, $y$
and $z$ directions, respectively. We assume the
spatial dependence of the pair potential as
$\Delta =\Delta _{0}$ $(zero)$ with $i\geq 1$ $\left( <1\right) $
for junctions along $x$-axis
and $\Delta =\Delta _{0}$ $(zero)$ with $n\geq 1$ $\left( <1\right)
$ for those along $z$-axis.

By applying the Bogoliubov transformation in the above
lattice Hamiltonian
\begin{equation}
c_{ijn\sigma }=\sum\nolimits_{\nu }u_{ijn\sigma }^{\nu }\hat{\gamma}_{\nu
}+v_{ijn\sigma }^{\nu \ast }\hat{\gamma}_{\nu }^{\dag },
\end{equation}%
we can obtain the lattice version of the BdG equations
\begin{equation}
\left\{
\begin{array}{c}
\varepsilon _{\nu }u_{ijn\uparrow }^{\nu }=(-tu_{i+1,jn\downarrow }^{\nu
}+tu_{i-1,jn\downarrow }^{\nu }+itu_{i,j+1,n\downarrow }^{\nu
}-itu_{i,j-1,n\downarrow }^{\nu }+t_{z}u_{ij,n+1\uparrow }^{\nu
}+t_{z}u_{ij,n-1\uparrow }^{\nu })/2+ \\
\left( 2m-t_{z}\cos Q-\mu \right) u_{ijn\uparrow }^{\nu
}-m(u_{i+1,jn\uparrow }^{\nu }+u_{i-1,jn\uparrow }^{\nu }+u_{i,j+1,n\uparrow
}^{\nu }+u_{i,j-1,n\uparrow }^{\nu })/2+\Delta v_{ijn\downarrow }^{\nu }, \\
\\
\varepsilon _{\nu }u_{ijn\downarrow }^{\nu }=(tu_{i+1,jn\uparrow }^{\nu
}-tu_{i-1,jn\uparrow }^{\nu }+itu_{i,j+1,n\uparrow }^{\nu
}-itu_{i,j-1,n\uparrow }^{\nu }-t_{z}u_{ij,n+1\downarrow }^{\nu
}-t_{z}u_{ij,n-1\downarrow }^{\nu })/2+ \\
\left( -2m+t_{z}\cos Q-\mu \right) u_{ijn\downarrow }^{\nu
}+m(u_{i+1,jn\downarrow }^{\nu }+u_{i-1,jn\downarrow }^{\nu
}+u_{i,j+1,n\downarrow }^{\nu }+u_{i,j-1,n\downarrow }^{\nu })/2-\Delta
v_{ijn\uparrow }^{\nu }, \\
\\
\varepsilon _{\nu }v_{ijn\uparrow }^{\nu }=(tv_{i+1,jn\downarrow }^{\nu
}-tv_{i-1,jn\downarrow }^{\nu }+itv_{i,j+1,n\downarrow }^{\nu
}-itv_{i,j-1,n\downarrow }^{\nu }-t_{z}v_{ij,n+1\uparrow }^{\nu
}-t_{z}v_{ij,n-1\uparrow }^{\nu })/2+ \\
\left( -2m+t_{z}\cos Q+\mu \right) v_{ijn\uparrow }^{\nu
}+m(v_{i+1,jn\uparrow }^{\nu }+v_{i-1,jn\uparrow }^{\nu }+v_{i,j+1,n\uparrow
}^{\nu }+v_{i,j-1,n\uparrow }^{\nu })/2-\Delta u_{ijn\downarrow }^{\nu }, \\
\\
\varepsilon _{\nu }v_{ijn\downarrow }^{\nu }=(-tv_{i+1,jn\uparrow }^{\nu
}+tv_{i-1,jn\uparrow }^{\nu }+itv_{i,j+1,n\uparrow }^{\nu
}-itv_{i,j-1,n\uparrow }^{\nu }+t_{z}v_{ij,n+1\downarrow }^{\nu
}+t_{z}v_{ij,n-1\downarrow }^{\nu })/2+ \\
\left( 2m-t_{z}\cos Q+\mu \right) v_{ijn\downarrow }^{\nu
}-m(v_{i+1,jn\downarrow }^{\nu }+v_{i-1,jn\downarrow }^{\nu
}+v_{i,j+1,n\downarrow }^{\nu }+v_{i,j-1,n\downarrow }^{\nu })/2+\Delta
u_{ijn\uparrow }^{\nu }.%
\end{array}%
\right.   \label{bdG}
\end{equation}%
Here, wave functions of the NS junction can be written as%
\begin{eqnarray}
\Psi _{\alpha }^{N}\left( \mathbf{r}\right)  &=&\sum\nolimits_{\beta }\left[
\xi _{\alpha }^{e}\left( \mathbf{r}\right) +r_{\alpha \beta }^{e}\xi _{\beta
}^{e}\left( \mathbf{r}\right) +r_{\alpha \beta }^{h}\xi _{\beta }^{h}\left(
\mathbf{r}\right) \right] , \\
\Psi _{\alpha }^{S}\left( \mathbf{r}\right)  &=&\sum\nolimits_{\delta }\left[
t_{\alpha \delta }^{e}\eta _{\delta }^{e}\left( \mathbf{r}\right) +t_{\alpha
\delta }^{h}\eta _{\delta }^{h}\left( \mathbf{r}\right) \right] ,
\end{eqnarray}%
where $\alpha $, $\beta $ and $\delta $ denote corresponding Weyl cones. The
spinors $\xi ^{e\left( h\right) }\left( \mathbf{r}\right) $ and $\eta
^{e\left( h\right) }\left( \mathbf{r}\right) $ can be solved by Eq.(\ref{bdG}%
):
\begin{equation}
\xi ^{e}(\mathbf{r})=\frac{1}{\sqrt{N_{e}}}\left(
\begin{array}{c}
t\left( i\sin k_{x}^{e}+\sin k_{y}^{e}\right)  \\
\mathcal{M}\left( k_{x}^{e},k_{y}^{e},k_{z}^{e}\right) -\mu -E \\
0 \\
0%
\end{array}%
\right) e^{i(k_{x}^{e}x+k_{y}^{e}y+k_{z}^{e}z)},
\end{equation}%
\begin{equation}
\xi ^{h}(\mathbf{r})=\frac{1}{\sqrt{N_{h}}}\left(
\begin{array}{c}
0 \\
0 \\
-t\left( i\sin k_{x}^{h}-\sin k_{y}^{h}\right)  \\
-\mathcal{M}\left( k_{x}^{h},k_{y}^{h},k_{z}^{h}\right) +\mu -E%
\end{array}%
\right) e^{i(k_{x}^{h}x+k_{y}^{h}y+k_{z}^{h}z)},
\end{equation}%
with
\begin{equation}
\mathcal{M}\left( k_{x},k_{y},k_{z}\right) =t_{z}\left( \cos k_{z}-\cos
Q\right) +m\left( 2-\cos k_{x}-\cos k_{y}\right) ,
\end{equation}%
and
\begin{equation}
\eta (\mathbf{r})=\frac{1}{\sqrt{N_{S}}}\left(
\begin{array}{c}
\left[ \mathcal{G}\left( p_{x},p_{y},p_{z}\right) -\left( \mu -E\right)
^{2}+\Delta_{0} ^{2}\right] \left( t\sin p_{x}-it\sin p_{y}\right)  \\
i\left[ \mathcal{G}\left( p_{x},p_{y},p_{z}\right) -\left( \mu -E\right) ^{2}%
\right] \left[ \mu +E-\mathcal{M}\left( p_{x},p_{y},p_{z}\right) \right] +%
\left[ -\mu +E+\mathcal{M}\left( p_{x},p_{y},p_{z}\right) \right] \Delta _{0}^{2}
\\
i\left\{ -\mu ^{2}+\left[ E-\mathcal{M}\left( p_{x},p_{y},p_{z}\right) %
\right] ^{2}-t^{2}\left( \sin ^{2}p_{x}+\sin ^{2}p_{y}\right) -\Delta
^{2}\right\} \Delta_{0}  \\
2\left( \mu -\mathcal{M}\left( p_{x},p_{y},p_{z}\right) \right) \left( t\sin
p_{x}-it\sin p_{y}\right) \Delta_{0}
\end{array}%
\right) e^{i(p_{x}x+p_{y}y+p_{z}z)},
\end{equation}%
with
\begin{equation}
\mathcal{G}\left( p_{x},p_{y},p_{z}\right) \mathcal{=M}^{2}\left(
p_{x},p_{y},p_{z}\right) +t^{2}\left( \sin ^{2}p_{x}+\sin ^{2}p_{y}\right).
\end{equation}%
Here, $N_{e}$, $N_{h}$, and $N_{S}$ are normalization constants. $k_{x}^{e\left(h\right) }$,
$k_{y}^{e\left( h\right) }$ and $k_{z}^{e\left( h\right) }$ satisfy
the dispersion relation%
\begin{equation}
E^{e}=\sqrt{\mathcal{G}\left( k_{x}^{e},k_{y}^{e},k_{z}^{e}\right) }-\mu ,
\end{equation}%
\begin{equation}
E^{h}=-\sqrt{\mathcal{G}\left( k_{x}^{h},k_{y}^{h},k_{z}^{h}\right) }+\mu ,
\end{equation}%
and $p_{x}$, $p_{y}$ and $p_{z}$ satisfy%
\begin{equation}
E^{S}=\sqrt{\left( \mu -\sqrt{\mathcal{G}\left( p_{x},p_{y},p_{z}\right)
+\left( \mathcal{M}\left( p_{x},p_{y},p_{z}\right) \Delta_{0} /\mu \right) ^{2}}%
\right) ^{2}+\Delta _{0}^{2}\left( 1-\mathcal{M}^{2}\left(
p_{x},p_{y},p_{z}\right) /\mu ^{2}\right) }.
\end{equation}%
Using the boundary condition\cite{S_Kroemer}:
\begin{eqnarray}
\hat{t}\Psi _{\alpha }^{N}\left( i=1\right)  &=&\chi \hat{t}\Psi _{\alpha
}^{S}\left( i=1\right) , \\
\chi \hat{t}^{\prime }\Psi _{\alpha }^{N}\left( i=0\right)  &=&\hat{t}%
^{\prime }\Psi _{\alpha }^{S}\left( i=0\right) ,
\end{eqnarray}%
for the junction along $x$-axis and
\begin{eqnarray}
\hat{t}_{z}\Psi _{\alpha }^{N}\left( n=1\right)  &=&\chi \hat{t}_{z}\Psi
_{\alpha }^{S}\left( n=1\right) , \\
\chi \hat{t}_{z}^{\prime }\Psi _{\alpha }^{N}\left( n=0\right)  &=&\hat{t}%
_{z}^{\prime }\Psi _{\alpha }^{S}\left( n=0\right) ,
\end{eqnarray}%
for the junction along $z$-axis, one can obtain the coefficients $r_{\alpha \beta
}^{e\left( h\right) }$ and $t_{\alpha \delta }^{e\left( h\right) }$.
Here, $\hat{t}$ and $\hat{t}_{z}$ represent the effective hopping term in BdG
equations given by%
\begin{eqnarray}
\hat{t} &=&\left( -it\sigma _{y}\tau _{z}-m\sigma _{z}\tau _{z}\right) /2, \\
\hat{t}^{\prime } &=&\left( it\sigma _{y}\tau _{z}-m\sigma _{z}\tau
_{z}\right) /2, \\
\hat{t}_{z} &=&\hat{t}_{z}^{\prime }=t_{z}\sigma _{z}\tau _{z}/2.
\end{eqnarray}%
We define $\chi $ to describe the transmissivity of NS
junction. $\chi =0$ ($\chi =1$) corresponds to the edge (perfect
transmitting junction). Finally, we obtain the charge current:%
\begin{equation}
I\left( V\right) =\eta _{1}\int dE\left[ f\left( E-eV\right) -f\left(
E\right) \right] \sigma _{S}\left( E\right) ,
\end{equation}%
where%
\begin{equation}
\sigma _{S}\left( E\right) =\eta _{2}\sum\nolimits_{\alpha ,\mathbf{k}%
_{\parallel }}\sigma _{S}^{\alpha }\left( E,\mathbf{k}_{\parallel }\right) ,
\end{equation}%
\begin{equation}
\sigma _{S}^{\alpha }\left( E,\mathbf{k}_{\parallel }\right)
=\sum\nolimits_{\beta }\mathrm{Re}[1-\frac{v_{\beta }^{\left( e\right)
}\left( E\right) }{v_{\alpha }^{\left( e\right) }\left( E\right) }\left\vert
r_{\alpha \beta }^{e}\right\vert ^{2}+\frac{v_{\beta }^{\left( h\right)
}\left( E\right) }{v_{\alpha }^{\left( e\right) }\left( E\right) }\left\vert
r_{\alpha \beta }^{h}\right\vert ^{2}].
\end{equation}%
The summation runs over all the indices of Weyl cones
$\alpha$ and $\mathbf{k}_{\parallel }$.
The quantity $\mathbf{k} _{\parallel }$ denotes $\left( k_{x},k_{y}\right) $
and $\left( k_{y},k_{z}\right)$
in the junction along $x$-axis and that along $z$-axis, respectively.

$v_{\beta }^{\left( e\right) }\left( E\right) $ are the group
velocities which can be derived from dispersion relation of the bulk energy spectrum $\left(
1/\hbar \right) \partial E/\partial k_{x}$.
$\eta _{1\left( 2\right) }$ is
the constant determined by the geometry of the microconstruction.
We calculate  normalized conductance%
\begin{equation}
\sigma _{n}\left( eV\right) =\sigma _{S}\left( eV\right) /\sigma _{N}\left(
eV\right) ,
\end{equation}%
where $\sigma _{N}\left( eV\right) $ is the conductance in the normal
state. It is noted that $\eta _{1\left( 2\right) }$ does not appear in the
expression of the normalized conductance $\sigma_{n}(eV)$.

\begin{figure}[htbp]
 \begin{center}
  \includegraphics*[width = 110 mm]{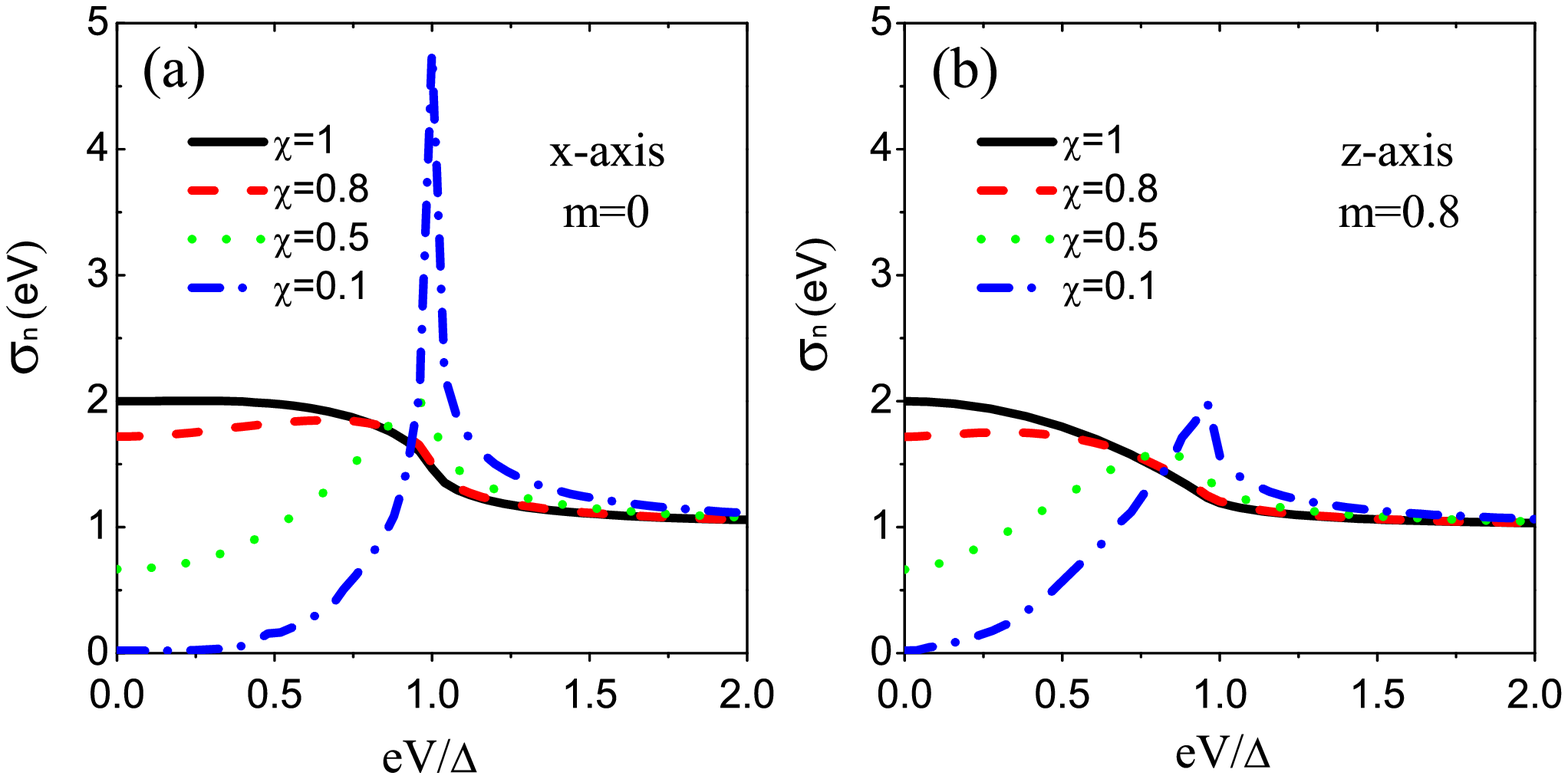}
 \end{center}
 \caption{(Color online) (a) Normalized tunneling conductance
 as a function of bias
voltage $\left( eV/\Delta \right) $ for junctions along
(a) $x$-axis with $m=0$ and (b)
$z$-axis with $m=0.8$.
 Other parameters are as the same as in Fig.2 in the main text. }
 \label{smfig2}
\end{figure}

In the main text, we have shown that the zero biased conductance peak (ZBCP)
emerges when the NS junction is along the $x$-direction.
But as shown in Fig.\ref{smfig2}
(a), the present ZBCP depends on $m$ and it vanishes for $m=0$.
This is consistent with the discussion
based on the winding number in main text since the
projected Fermi surfaces are overlapped and
resulting surface Andreev
bound states are absent for $m=0$.
In the junction along $z$-axis, the resulting $\sigma _{n}$ is
insensitive to $m$. Since there is no surface Andreev bound states along
this direction, ZBCP does not appear in the limit of low transmissivity
$\chi \rightarrow 0$. The line shapes of $\sigma _{n}$  are the
essentially the same as those of the bulk density of states as shown in Fig.\ref{smfig2}
(b).

\end{document}